\documentstyle[12pt]{article}
\advance\hoffset -0.5in
\advance\voffset -0.6in
\def\pindent{$\mbox{}$\indent}  

\def\pindent{$\mbox{}$\indent}  

\def\bbt#1{\bibitem{#1} \label{#1}}
\def\bbr#1#2{[\ref{#1}--\ref{#2}]}

\def\eqnroot{\arabic{equation}}
\def\defaultnumbering{\renewcommand{\theequation}{\eqnroot}}
\defaultnumbering
\newcounter{sen}

\def\eanumbering{\renewcommand{\theequation}{\eqnroot\alph{sen}}}
\def\ealabel#1{%
      \newcounter{#1eqns}\setcounter{#1eqns}{\arabic{equation}}          }
\def\earef#1{
      \arabic{#1eqns}}

\def\be{\begin{equation}}
\def\ee{\end{equation}}
\def\ba{\begin{eqnarray}}
\def\ea{\end{eqnarray}}
\def\bal{\eanumbering \setcounter{sen}{1}\begin{eqnarray}}
\def\alp{\\ \addtocounter{equation}{-1}\stepcounter{sen}}
\def\eal{\end{eqnarray}\defaultnumbering\noindent}
\def\ban{\begin{eqnarray*}}
\def\ean{\end{eqnarray*}}

\def\th#1{$#1$\/th}
\def\qal{\langle\!\langle}
\def\qar{\rangle\!\rangle}
\def\bqal{\biggl\langle\!\!\biggl\langle}
\def\bqar{\biggr\rangle\!\!\biggr\rangle}

\def\bxi{\mbox{\boldmath $\xi$}}

\def\bs{{\bf s}}
\def\erf{\mbox{erf}}
\def\ptl#1{\frac{\partial}{\partial #1}}
\def\sfrac#1#2{{\textstyle \frac{#1}{#2}}}
\def\ds#1{{\displaystyle #1}}

\def\lsim{\mbox{{\raisebox{-1mm}{$\stackrel{\textstyle <}{\sim}$}}}}

\def\str#1#2{\mbox{{\raisebox{-2.5mm}{$\stackrel{\mbox{#1}}{\scriptstyle
#2}$}}}}

\def\ie{i.e., }
\def\etal{{\it et al.}}

\def\el#1{Europhys.~Lett.~{\bf #1}}

\def\jpa#1{J.~Phys.~A~{\bf #1}}

\def\jpp#1{J.~Phys.~(Paris)~{\bf #1}}

\def\pra#1{Phys.~Rev.~A~{\bf #1}}

\def\prl#1{Phys.~Rev.~Lett.~{\bf #1}}

\def\zpb#1{Z.~Phys.~B~{\bf #1}}

\begin{document}

\thispagestyle{empty}

\vskip -0.6 in
\hfill \mbox{\small\tt SNUTP 93-26}
\vskip 0.6 in
\centerline{\Large \bf Optimal storage capacity of neural networks }
\vskip 0.05in
\centerline{\Large \bf at finite temperatures }
\vskip 0.3 in
\centerline{G. M. Shim$\ ^\dagger $, D. Kim$\ ^{\dagger \dagger}$,
         and M. Y. Choi$\ ^{\dagger \dagger}$ }
\vskip 0.1 in
\centerline{\it Department of Physics and Center for Theoretical Physics}
\centerline{\it Seoul National University, Seoul 151-742, Korea}

\vspace{0.4in}
\begin{quotation}  

Gardner's analysis of the optimal storage capacity of neural networks is
extended to study finite-temperature effects.
The typical volume of the space of interactions is calculated for
strongly-diluted networks as a function of the storage ratio $\alpha$,
temperature $T$, and the tolerance parameter $m$, from which the
optimal storage capacity $\alpha_c$ is obtained as a function of $T$ and $m$.
At zero temperature it is found that $\alpha_c = 2$ regardless of $m$ while
$\alpha_c$ in general increases with the tolerance at finite temperatures.
We show how the best performance for given $\alpha$ and $T$ is obtained, which
reveals a first-order transition from high-quality performance to low-quality
one at low temperatures. An approximate criterion for recalling, which is valid
near $m=1$, is also discussed.

\vskip 0.2in
\footnoterule
\noindent  $\ ^\dagger $  Present address: Institute voor Theoretische Fysica,
Katholieke Universiteit Leuven, B--3001 Leuven, Belgium. e-mail:
fgbda28@cc1.kuleuven.ac.be \\
 $\ ^{\dagger \dagger} $  e-mails: dkim@phya.snu.ac.kr, mychoi@phya.snu.ac.kr
\\

\end{quotation}


\newpage

\section{Introduction}
                      \pindent
Recently, the tools of statistical mechanics have been extensively applied to
the study of collective properties of neural networks~\cite{rev89};
spin glass theory has played an important role in the growth of this new
field~\cite{Mezard86}.
In particular, the optimal (error-free) storage capacity for recurrent networks
can be obtained by
calculating the typical fractional volume of the space of interactions
$(\{J_{ij}\})$ satisfying the condition that for a given set of patterns
each pattern is a fixed point of the deterministic (zero-temperature) dynamics
\be
      s_i(t+1) = \mbox{sign}[\sum_j J_{ij}s_j(t)]\,,  \label{eq:det}
\ee
where $s_i(t)\;(=\pm1)\;(i=1,\ldots,N)$ represents the state of the \th{i}
neuron at
time $t$, and the synaptic coupling $J_{ij}$ determines the contribution of a
signal fired by the \th{j} neuron to the action potential on the \th{i} neuron.
This approach of systematic exploration of the space of interactions, which
was pioneered by Gardner~\cite{Gardner87} and reformulated in terms of
canonical ensemble calculation~\cite{Gardner88},
has been applied in various directions~\bbr{Gardner88}{Kanter92}.
The Hopfield model with general continuous couplings has been found to be
capable of storing at most two uncorrelated random patterns per neuron
without errors and
larger number of patterns for biased patterns~\cite{Gardner87}.
The network with discrete (Ising-type) couplings has been also extensively
investigated since the replica-symmetry theory was reported to yield
wrong results for the optimal storage
capacity~\cite{Gardner88,Krauth89,Gutfreund90}.
The method is not limited to Hopfield-type neural networks but applicable to
multilayer networks as well as to simple
perceptrons~\cite{Gardner89a,Kanter92}.
However, Gardner's method is based on the concept of fixed points of the
dynamics, and obviously does not work if the dynamics is stochastic (\ie at
finite temperatures). In addition, it requires perfect matching so that each
pattern is inerrably recalled at every site whereas in practice one usually
considers a neural network to be {\it remembering\/} or {\it recalling\/}
if the overlap between the network state and one of the patterns is larger
than some given value.

In this paper, we propose a scheme to define the optimal storage
capacity at finite temperatures and study its temperature dependence.
We introduce the tolerance parameter $m(\leq 1)$ in such a way that the
$m\rightarrow 1$ limit corresponds to the perfect recall while $(1-m)/2$
measures the error allowed. We then calculate the typical
fractional volume of the space of interactions for
extremely-diluted networks as a function of the storage ratio $\alpha$,
temperature $T$, and the tolerance parameter $m$, which leads to the
optimal storage capacity $\alpha_c$ as a function of $T$ and $m$.
At zero temperature it is found that $\alpha_c=2$ regardless of the
tolerance parameter $m$.
At finite temperatures, on the other hand, the optimal storage capacity
vanishes  in the perfect
matching limit ($m\rightarrow1$) and in general increases with the tolerance.
We then discuss how the best performance is obtained for given $\alpha$ and
$T$.
We also propose an alternative criterion for recalling, which may be regarded
as
a simple approximate scheme to define the optimal storage capacity, and
consider
the optimal storage capacity of dynamic model~\cite{Choi88} as well as of the
extremely-diluted model in this approximate scheme.

The contents of this paper are as follows: In Sec.~II, we
propose a scheme to define the optimal storage capacity at finite temperatures
together with an approximate scheme. Section~III is devoted to the
calculation of the optimal storage capacity of
an extremely-diluted neural network
while Sec.~IV presents results of the approximate scheme.
A brief discussion is given in Sec.~V.


\section{Optimal storage capacity at finite temperatures}
               \pindent
One usually takes into account internal noise in the functioning of
a neuron by extending the deterministic evolution rule~(\ref{eq:det})
to a stochastic one\/:
\be
    P[s_i(t+1)=\pm1]=\sfrac12\{1\pm\tanh[\beta\sum_j J_{ij}s_j(t)]\}\,,
       \label{eq:stoc}
\ee
where the inverse {\it temperature\/} ($\beta\equiv 1/T$) measures the width of
the threshold region, i.e., the level of synaptic noise.
The state $\bs\equiv\{s_i\}$ of the network of $N$ neurons evolves
stochastically according to Eq.~(\ref{eq:stoc}).
A given set of states of the network to be memorized by adjusting
appropriately the couplings is called the set of  patterns.
We now define the overlap $M_\mu(t)$ between the network state and the \th{\mu}
pattern $\bxi^\mu\equiv\{\xi_i^\mu\}\;(\mu=1,\ldots,p)$ by
$$
    M_\mu(t) \equiv \frac1N\sum_{i=1}^N\xi_i^\mu s_i(t)\,,
$$
which also evolves stochastically along with $\bs(t)$.
When a network is recalling pattern $\mu$, the time average of $M_\mu(t)$
over the time scale sufficiently longer than the observational time
but shorter than the life time of the local energy minimum
should be close to unity.
However, since the dynamics is stochastic, it cannot be strictly unity as in
the zero-temperature dynamics. Therefore, we introduce the tolerance parameter
$m$ in such a way that the network is considered to be remembering the \th{\mu}
pattern if
$$
    \overline{M}_\mu  \equiv \frac1{N_t}\sum_{t=1}^{N_t}M_\mu(t)\: > m
$$
with $N_t$ in the appropriate range
as mentioned above.
The quantity $(1-m)/2$ is the maximum
error allowed for the network to be qualified as {\it functioning}.
It is expected that the time average $ \overline{M}_\mu$ is
equivalent to the restricted thermal average
$$
       \langle M_\mu \rangle \equiv \frac1N\sum_{i=1}^N
          \xi_i^\mu \langle s_i\rangle\,,
$$
where the thermal measure is restricted within a single pure state
(containing the configuration $\bxi^\mu$)~\cite{Mezard89}.
In the stationary state the activity $\langle s_i \rangle $
of the \th{i} neuron is determined by the coupled equations
$$
     \langle s_i\rangle = \langle \tanh(\beta\sum_j J_{ij}
      s_j) \rangle = \tanh(\beta\sum_j J_{ij} \langle
      s_j \rangle)\,,
$$
where a mean-field approximation has been used. Such an approximation is
expected to be valid for diluted networks which we mainly consider in this
work. Otherwise, a reaction term may be  necessary.
The optimal storage capacity is given by the upper bound of the storage ratio
$\alpha\equiv p/N$, where $p$ is the number of stored patterns.
The problem reduces, according to Gardner~\cite{Gardner87}, to the calculation
of the typical fractional volume of the space of interactions which satisfies
the following conditions\/:
\be
   \frac1N\sum_{i=1}^N\xi_i^\mu\tanh\biggl(\frac{\beta}{\sqrt{N}}\sum_j
     J_{ij} \langle s_j \rangle\biggr)\; > m\,, \label{eq:stfpe}
\ee
and
\be
   \sum_j(J_{ij})^2 = N \hspace{3em}\mbox{for each $i$}\,.  \label{eq:normal}
\ee
The condition in Eq.~(\ref{eq:normal}) is required to fix the scale of
temperature $T$.
The optimal storage capacity at temperature $T$ is then determined by vanishing
of this fractional volume, which leads to $\alpha_c$ as a function of $T$ and
$m$.
Application of the above scheme to an extremely-diluted neural network will
be given in the following section.
In this model, only an extremely small fraction of couplings among
neurons are connected so that its dynamics can be solved in a rather simple
manner~\cite{Derrida87,Gardner89c}.  However, calculation of the fractional
volume for other generic neural network is formidable due to the thermal
average to be performed within one  pure state.
As a simple attempt, one may use the approximation $\langle s_i
\rangle \approx \xi_i^\mu$ and  replace Eq.~(\ref{eq:stfpe}) by
\be
  \frac1N\sum_{i=1}^N\xi_i^\mu\tanh\biggl(\frac{\beta}{\sqrt{N}}\sum_jJ_{ij}
          \xi_j^\mu \biggr)\; > m\,, \label{eq:appfpe}
\ee
which states that the network state evolved by one time step from a given
pattern has overlap with that pattern greater than $m$.
This simple criterion presumably leads to results similar to those of
Eq.~(\ref{eq:stfpe}) for $m$ close to unity, where the network is expected
to hover around the configuration $\bxi^\mu$ during the recalling state.
The validity of this approximate scheme with regard to diluted networks
is discussed in Sec.~IV.


\section{Extremely-diluted neural network}
           \pindent
In this section the proposed scheme is applied to an extremely-diluted neural
network,
where, on the average, there are $C\;(\lsim\log N)$ connections per neuron.
Such a model was first studied by Derrida \etal~\cite{Derrida87}, and later its
properties
of the basin of attraction was studied by Gardner~\cite{Gardner89c} and
by Amit \etal~\cite{Amit90}.
The reason why we can implement the scheme exactly is that the dynamics of the
network can be solved in a rather simple manner~\cite{Derrida87,Gardner89c}.
Extending the method of  Keppler and Abbott~\cite{Keppler88},
we describe the time evolution of
the overlap $M_\mu(t)$ between pattern $\bxi^\mu$ and the network state
by the one-step recursion relation
\be
     M_\mu(t+1) = F_{{\bf h}^\mu}[M_\mu(t)]\,. \label{eq:osrec}
\ee
Here the {\it map\/} $F_{{\bf h}^\mu}(x)$ is defined by
$$
        F_{{\bf h}^\mu}(x) \equiv \frac{1}{N}\sum_{i=1}^N\int Dz\,\tanh[
                  \beta(\sqrt{1-x^2}\:z+h_i^\mu x)]\,,
$$
where $\beta\:(\equiv T^{-1})$ is the inverse temperature, $Dz$ denotes the
Gaussian measure: $Dz\equiv\exp(-z^2/2)\,dz/\sqrt{2\pi}$, and
the subscript ${\bf h}^\mu$ denotes the $\{h_i^\mu\}$-dependence of the map
with
\be
        h_i^\mu \equiv \xi_i^\mu\sum_j\frac{J_{ij}}{\sqrt{C}}\xi_j^\mu\,.
          \label{eq:pot}
\ee
In the stationary state, $M_\mu(t)$ approaches $M_\mu(t\rightarrow\infty)=
M_\mu^\ast$ of which value is determined by the stable
fixed point solution $x$ satisfying
\bal
          x &=& F_{{\bf h}^\mu}(x)  \label{eq:ssl} \alp \ealabel{map}
          |\ptl{x}F_{{\bf h}^\mu}(x)|&\equiv&|F_{{\bf h}^\mu}'(x)|\,<1\,,
	            \label{eq:sstb}
\eal
where Eq.~(\ref{eq:sstb}) has been imposed to guarantee its stability.
For given tolerance parameter $m$, the network is considered to be remembering
the \th{\mu} pattern if the value of the stationary overlap $M_\mu^\ast$ is
greater than $m$.

Now the main quantity to calculate is the fractional volume of the space of
interactions $(\{J_{ij}\})$ for which every pattern can be remembered.
The normalization condition is now given by
$$
         \sum_j (J_{ij})^2 = C
$$
instead of Eq.~(\ref{eq:normal}).
 The number of  the solutions of Eqs.~(\earef{map})
 with its value greater than $m$ is formally given by
\be
   {\cal N}_{{\bf h}^\mu} \equiv \int_m^1dM\,\delta(M-F_{{\bf h}^\mu}(M))
          \:|1-F_{{\bf h}^\mu}'(M)|\:\theta(1-|F_{{\bf h}^\mu}'(M)|)
	  \label{eq:numsol}
\ee
so that the fractional volume can be written as
$$
   V_0 = \frac{\ds{ \int[\prod_{i\ne j}}dJ_{ij}]\ds{\prod_i}\delta\biggl(
       \ds{\sum_j}(J_{ij})^2-C \biggr)\ds{\prod_{\mu=1}^{\alpha C}}
        \theta({\cal N}_{{\bf h}^\mu}) }{
          \ds{\int[\prod_{i\ne j}} dJ_{ij}]
          \ds{\prod_i} \delta\biggl(\ds{\sum_j}(J_{ij})^2-C\biggr)}
$$
where $\theta(x)$ is the step function and the number of stored patterns has
been scaled according to $p\equiv\alpha C$.
Here we are mainly interested in obtaining the optimal storage capacity
$\alpha_c$. If the number of stored patterns exceeds $\alpha_c C$, there is no
typical network $(\{J_{ij}\})$ that yields the value of the stationary overlap
greater than $m$ for all patterns.
In the limit $\alpha \rightarrow\alpha_c$, the number ${\cal N}_{{\bf h}^\mu}$
of stable solutions approaches zero, and we may replace the step function
$\theta({\cal N}_{{\bf h}^\mu})$ by ${\cal N}_{{\bf h}^\mu}$.
Furthermore the fractional volume vanishes only if ${\cal N}_{{\bf h}^\mu}
= 0$ for some $\mu$, which implies that the replacement $\theta(x)\rightarrow
x$
would not affect the optimal storage capacity.
The fractional volume to calculate is now given by
\be
   V = \frac{\ds{ \int[\prod_{i\ne j}}dJ_{ij}]\ds{\prod_i}\delta\biggl(
       \ds{\sum_j}(J_{ij})^2-C \biggr)\ds{\prod_{\mu}}{\cal N}_{{\bf h}^\mu}}{
          \ds{\int[\prod_{i\ne j}} dJ_{ij}]
          \ds{\prod_i} \delta\biggl(\ds{\sum_j}(J_{ij})^2-C\biggr)}\,.
\ee
Replacement of $\theta (x)$ by $x$ in general is also justified in the
following sense.  We may assume that, for typical $\{\bxi^\mu\}$,
${\cal N}_{{\bf h}^\mu}$ possesses finite, system size independent upper
bound almost everywhere in the interaction space. This is reasonable
since the map $F_{{\bf h}^\mu} (x)$ is an average of $N$ functions of the
form $$ \int Dz\,\tanh[
                  \beta(\sqrt{1-x^2}\:z+h x)] \, .$$
This function is smooth and monotonic in $x$ with derivative having the same
sign as $h$. Therefore the average  over many possible $h$ is a sum of
two parts: the
monotonically increasing part from contributions of $h>0$ and
the monotonically decreasing part from $h<0$. So in practice there are only a
few solutions at most.
 If we denote the upper bound by ${\cal N}_0$,
we then have the identity
$$ \theta ( {\cal N}_{{\bf h}^\mu} ) \leq {\cal N}_{{\bf h}^\mu} \leq
   {\cal N}_0 \, \theta ( {\cal N}_{{\bf h}^\mu} ) \, . $$
   Integrating this over the interaction space, we immediately see that
   $$  V_0 \leq V \leq {\cal N}_0^{\alpha C} \, V_0  \, . $$
   However, the fractional volumes are of the order of $\exp (- C N)$ so that
   $(\log V)/C N$  is the same as $(\log V_0)/C N$ in the thermodynamic limit
   $N \rightarrow \infty$.

In this work, we consider the case that every pattern $\xi_i^\mu$ to be
stored is an independently distributed random variable, taking the value $\pm1$
with equal probabilities.
The typical fractional volume $\overline{V}\equiv\exp(\qal\log V\qar)$ for the
random patterns involves averaging of $\log V$ over the distribution of the
random patterns $\{\bxi^\mu\}$, which may be obtained through the use of
 the well-known replica trick.
To facilitate the averaging over the distribution of the random patterns,
 we introduce $\delta$-functions describing Eq.~(\ref{eq:pot}) with the
help of the conjugate variable $\hat{h}_i^\mu$ raised to the exponential form
$$
      \delta(h_i^\mu-\xi_i^\mu\sum_j\frac{J_{ij}}{\sqrt{C}}\xi_j^\mu)
          = \int\frac{d\hat{h}_i^\mu}{2\pi}\exp[i\hat{h}_i^\mu(
        h_i^\mu-\xi_i^\mu\sum_j\frac{J_{ij}}{\sqrt{C}}\xi_j^\mu)]\,.
$$
The average over the random patterns for the replicated volume $\qal V^n\qar$
affects the exponential factor containing $\xi_i^\mu$ in the above expression,
and leads to the following:
\ban
   && \hspace*{-3em}\bqal\prod_{\alpha=1}^n
     \exp\biggl(-i\sum_{\mu i}\hat{h}_i^{\mu\alpha} \xi_i^\mu
       \sum_j\frac{J_{ij}^\alpha}{\sqrt{C}}\xi_j^\mu \biggr) \bqar_{
             \{\bxi^\mu\}} \\ && =
      \prod_\mu \exp\biggl(-\sfrac12\sum_{\alpha\beta i}\hat{h}_i^{\mu\alpha}
      \hat{h}_i^{\mu\beta}\sum_j\frac{J_{ij}^\alpha J_{ij}^\beta}{C}
      -\sfrac12\sum_{\alpha i}\hat{h}_i^{\mu\alpha} \sum_{\beta j}
       \hat{h}_j^{\mu\beta}\frac{J_{ij}^\alpha J_{ji}^\beta}{C} \biggr)\,,
\ean
where $\alpha$ and $\beta$ are the replica indices and,
for an extremely-diluted network, the cumulant expansion has been cut-off
at the second order~\cite{Gardner89b}.
Following Ref.~\cite{Gardner89b},
we assume the second term to be independent of site $i$, so that
$$
  \sum_{\beta j} \hat{h}_j^{\mu\beta}\frac{J_{ij}^\alpha J_{ji}^\beta}{C}
  = \frac1N \sum_i \sum_{\beta j} \hat{h}_j^{\mu\beta}\frac{J_{ji}^\beta
  J_{ij}^\alpha}{C}\,.
$$
Introducing the local order parameters
\ban
      q_{\alpha \beta}^i &\equiv& \frac1C\sum_j J_{ij}^\alpha J_{ij}^\beta
                                   \hspace{2em}  \\
      r_{\alpha \beta}^i &\equiv& \frac1C\sum_j J_{ij}^\alpha J_{ji}^\beta
\ean
together with their respective conjugate variables $Q_{\alpha \beta}^i$ and
$R_{\alpha \beta}^i$, we obtain
$$
 \qal V^n\qar = \frac1{\exp(nCN/2)} \int \prod_{\alpha i}\frac{
       dE_i^\alpha}{4\pi i}\int \prod_{\alpha<\beta\,i}
       \frac{dQ_{\alpha \beta}^i dq_{\alpha \beta}^i}{2\pi i/C}
     \int \prod_{\alpha\beta i}\frac{dR_{\alpha \beta}^i dr_{\alpha \beta}^i}{
     4\pi i/C}\,\exp(CG)\,,
$$
where
$$
   G \equiv \sfrac12\sum_{\alpha i}E_i^\alpha+\sum_{\alpha<\beta\,i}Q_{\alpha
     \beta}^i q_{\alpha \beta}^i+\sfrac12\sum_{\alpha\beta i}R_{\alpha\beta}^i
     r_{\alpha\beta}^i+G_1+\alpha G_2
$$
with $G_1$ and $G_2$ given by
\ban
   \exp(CG_1) &\equiv& \int \biggl[\prod_{\alpha\,i\ne j}dJ_{ij}^\alpha\biggr]
        \exp\biggl(-\sfrac12\sum_{\alpha ij}E_i^\alpha(J_{ij}^\alpha)^2
         -\sum_{\alpha<\beta}Q_{\alpha\beta}^iJ_{ij}^\alpha J_{ij}^\beta
         -\sfrac12\sum_{\alpha\beta ij}R_{\alpha\beta}^iJ_{ij}^\alpha
           J_{ji}^\beta \biggr)\,, \\
  \exp(G_2) &\equiv&\int\biggl[\prod_{\alpha i}\frac{d\hat{h}_i^\alpha
dh_i^\alpha}{
        2\pi}\biggr]\,[\prod_\alpha{\cal N}_{{\bf h}^\alpha}]\,
        \exp\biggl( i\sum_{\alpha i}\hat{h}_i^\alpha h_i^\alpha-\sfrac12
        \sum_{\alpha<\beta\,i}\hat{h}_i^\alpha \hat{h}_i^\beta
q_{\alpha\beta}^i
        \\ && \hspace*{1cm}
   -\frac1{2N}\sum_{\alpha\beta ij}\hat{h}_i^\alpha \hat{h}_j^\beta
r_{\alpha\beta}^j
       -\sfrac12\sum_{\alpha i}(\hat{h}_i^\alpha)^2\biggr)\,.
\ean
In the thermodynamic limit $(N\rightarrow\infty)$, $C$ also
approaches infinity, albeit slowly, and $\qal V^n \qar$ can be computed
through the use of the steepest-descent method.
In order to find the saddle point, we assume the replica- and site-symmetric
ansatz
\ban
     E_i^\alpha=E \hspace{2em} R_{\alpha\alpha}^i=S \hspace{2em}
     r_{\alpha\alpha}^i=s &&\\
     Q_{\alpha\beta}^i=Q \hspace{2em} q_{\alpha\beta}^i=q \hspace{2em}
     R_{\alpha\beta}^i=R &&\hspace{2em} r_{\alpha \beta}^i = r \hspace{2em}
     (\alpha \ne \beta)\,.
\ean
With this ansatz, the function $G$ in the limit $n\rightarrow 0$ takes the form
$$
    G = nN[\sfrac12(E-qQ+sS-rR)+g_1+\alpha g_2]\,,
$$
where
\ban
    g_1 &\equiv& -\sfrac14\biggl(\frac{Q+R}{E-Q+S-R}+ \frac{Q-R}{E-Q-S+R}
                    + \log(E-Q+S-R) \\ && + \log(E-Q-S+R)\biggr)\,, \\
    g_2 &\equiv& \frac1N\int\prod_{i=1}^NDt_i\;\log\biggl[
              \int\frac{dt_0}{\sqrt{2\pi/N}}\exp(-\sfrac12Nt_0^2)
            \int\prod_{i=1}^NDh_i\;{\cal N}_{\bf H}\,.
\ean
In the above expression ${\cal N}_{\bf H}$ is given by Eq.~(\ref{eq:numsol})
with ${\bf h}^\mu$ replaced by ${\bf H}\equiv\{H_i\}$, where
$H_i\equiv\sqrt{1-q}\,h_i-\sqrt{s-r}\,t_0-\sqrt{q}\,t_i$.
Since the saddle-point equations for the variables $E,\;S,\;Q$, and $R$
are algebraic, we can eliminate these variables and finally write the typical
fractional volume in the form
$$
 \overline{V} = \exp\biggl(CN\biggl[\sfrac12\log(1-q)+\sfrac14\log(1-x^2)
    +\sfrac12 \frac{q-rx}{(1-q)(1-x^2)}+\alpha g_2\biggr]\biggr)\,,
$$
where $x \equiv (s-r)/(1-q)$.

To manipulate $g_2$, we introduce the variable
$$
        \overline{M} \equiv \ptl{M}F_{\bf H}(M)
$$
together with its conjugate variable $\overline{\lambda}$ and use the integral
representation of $\delta$-function.  Noting the  range of the variable
$\overline{M}$, we obtain
\ban
    g_2 &=& \frac1N\int_{-\infty}^\infty\prod_{i=1}^NDt_i\;\log\biggl[
              \int_{-\infty}^\infty\frac{dt_0}{\sqrt{2\pi/N}}\int_m^1dM
            \int_{-i\infty}^{i\infty}
             \frac{d\lambda}{2\pi i/N}\int_{-1}^1d\overline{M}
             \int_{-i\infty}^{i\infty} \frac{d\overline{\lambda}}{2\pi i/N}
             (1-\overline{M}) \\
         &&\times\exp[-N(\sfrac12t_0^2+\lambda M+\overline{\lambda}
          \,\overline{M})]
         \prod_{i=1}^N\int Dh_i \exp[\lambda f(H_i,M)+\overline{\lambda}
          \overline{M})]\, \partial_Mf(H_i,M)]
             \biggr] \,,
\ean
where the functions $f(H,M)$ and $\partial_Mf(H,M)$ are given by
\ban
      f(H,M) &\equiv& \int Dz\,\tanh[\beta(\sqrt{1-M^2}\,z+MH)] \\
      \partial_Mf(H,M) &\equiv& \ptl{M}f(H,M)\,.
\ean
Now the integration over $\overline{M}$ is easily performed, and
in the thermodynamic limit, the steepest-descent method yields
\be
    g_2 = \str{max}{m\le M \le 1}\,\str{max}{t_0}\,\str{min}{\lambda,
           \overline{\lambda}} \biggl(-\sfrac12t_0^2-\lambda M+
            |\overline{\lambda}|+\int Dt\,\log\int Dh\,
            \exp[\lambda f(H,M)+\overline{\lambda}\,\partial_Mf(H,M)]\biggr)\,,
           \label{eq:g2}
\ee
where $H \equiv \sqrt{1-q}\,h-\sqrt{(1-q)x}\,t_0-\sqrt{q}\,t$.
In Eq.~(\ref{eq:g2}), we should take the minimum over $\lambda$ and
$\overline{\lambda}$ rather than the maximum because the integration
over $\lambda$ and $\overline{\lambda}$  runs along the imaginary axis
in the complex plane,
which should be deformed to pass the saddle point. When the saddle point
happens
to lie on the  real axis, one may conveniently sweep along the real axis and
the saddle point corresponds to the minimum point along the real axis.

Since $\overline{V}$ depends on $s$ only through $x$, it is straightforward to
show that $\overline{V}$ reaches its maximum at $x=0$ and it follows that we
can
set $t_0=0$ in Eq.~(\ref{eq:g2}).
Since $q$ represents the typical correlations of the solution of
Eqs.~(\earef{map}),
 the typical fractional volume should shrinks to zero as
$q$ approaches unity. Accordingly, the optimal storage capacity is then
determined in this limit. When $q$ approaches unity the last term in
Eq.~(\ref{eq:g2}) diverges as $\sim(1-q)^{-1}$, and we write
$$
  \int Dt\, \log\int Dh\,\exp[\lambda f(H,M)+\overline{\lambda}\,
      \partial_Mf(H,M)]
    \longrightarrow      \frac1{1-q}\int Dt\,\Omega_t(H_t)\,,
$$
where the function $\Omega_t(H)$ is then given by
$$
       \Omega_t(H) \equiv -\sfrac12(H+t)^2+\lambda(1-q) f(H,M)+
                          \overline{\lambda}(1-q)\,\partial_Mf(H,M)
$$
and $H_t$ is the value of $H$ leading to the maximum of $\Omega_t$.
Therefore $g_2$ also exhibits $(1-q)^{-1}$ divergence:
$$
   g_2 = \frac1{1-q}\,\str{max}{m\le M\le1}\,\str{min}{ \lambda,\overline{
         \lambda}}\biggl(-\lambda(1-q)M+|\overline{\lambda}(1-q)|+\int Dt\,
        \Omega_t( H_t)\biggr)\,,
$$
and the saddle-point equation over $\lambda$ reads
\be
     M = \int Dt\,f(H_t,M) \label{eq:min1}\,,
\ee
where the dependence on the variables $\lambda$ and $\overline{\lambda}$
is implicit through $H_t(\lambda,\overline{\lambda},M;T)$.
For the minimization over $\overline{\lambda}$,
one should consider two cases. The first case is that the minimum occurs at
$\overline{\lambda}=0$. This happens when the absolute value of $\int Dt\,
\partial_Mf(H_t,M)$ with $\overline{\lambda}=0$ and
$\lambda=\lambda_0$, where $\lambda_0$ is given by the solution of
Eq.~(\ref{eq:min1}) with $\overline{\lambda}=0$, is less than unity.
In the other case, the minimum occurs at $\overline{\lambda}\ne0$
and the saddle point in the $(\lambda,\overline{\lambda})$ plane is given
by the solution of Eq.~(\ref{eq:min1}) together with the equation
$$
    -\mbox{sign}(\overline{\lambda}) = \int Dt\,\partial_Mf(H_t,M)\,.
$$
In both cases, we denote the saddle point to be $(\lambda_0,\overline{
\lambda}_0)$, and write $g_2$ in the form
$$
 g_2 = -\frac1{2(1-q)}\,\str{min}{m\le M\le1} \alpha_0^{-1}(M;T)\,,
$$
where
\be
   \alpha_0^{-1}(M;T)\equiv \int Dt\:[t+H_t(\lambda_0,\overline{\lambda}_0,
   M;T)]^2\,.  \label{eq:mscM}
\ee
Combining the above, we finally obtain the typical fractional volume:
$$
   \overline{V} = \exp\biggl(\frac{CN}{2(1-q)}[1-\alpha\,\str{min
                }{m\le M\le1}\alpha_0^{-1}(M;T)]\biggr)\,,
$$
which vanishes for
\be
   \alpha > \alpha_c\equiv \str{max}{m\le M\le1}\alpha_0(M;T)\,.
    \label{eq:oscm}
\ee
Interestingly, $\alpha_0(M;T)$ also represents the maximum storage capacity
for the stationary value of the overlap $M_\mu^\ast$ being in the range $M\le
M_\mu^\ast \le M+\delta M$. Due to the mean-field nature of the network, the
optimal storage capacity is given by the maximum value of $\alpha_0$ for the
given range of $M$.

Since $\alpha_0(M;T)$ defined in Eq.~(\ref{eq:mscM}) involves minimization with
respect to two variables $(\lambda,\overline{\lambda})$
in addition to the two Gaussian
integrals over $t$ and $z$ (representing thermal average), we computed them
numerically.  Figure~\ref{fig:a0M}  shows  typical behavior of
$\alpha_0(M;T)$ for several values of $T$, with detailed behavior
for $T=0.3$ and 0.7 displayed in the inset.  There exist three types of
$M$-dependence of $\alpha_0(M;T)$ according to $T$.
When $T$ is higher than $T_1(\approx0.566)$, the maximum capacity $\alpha_0$
decreases monotonically with $M$. For $T<T_1$, $\alpha_0$ exhibits a local
minimum as well as a local maximum (as shown in the inset of
Fig.~\ref{fig:a0M}).
This local maximum (at nonzero $M$) is in fact the global maximum of
$\alpha_0$ for $T$ lower than $T_2(\approx0.414)$ whereas $\alpha_0$ reaches
its maximum at $M=0$ for $T_2<T<T_1$.

In contrast to the naive expectation, the maximum storage  capacity $\alpha_0$
is not monotonic with $M$ (or with the error allowed) when the temperature is
lower than $T_1$. It is of interest to note that
there are two kinds of fluctuations in the dynamics:
One is thermal fluctuations associated with the synaptic noise  and
controlled by the temperature $T$ while the other is {\it dynamical
fluctuations\/} described by $\sqrt{1-M^2}$. The latter fluctuations come from
the distribution of states with definite overlap $M$ and may be considered
to be driven by the dynamics itself.
In general thermal fluctuations randomize spin orientations and tend to
decrease the capacity whereas dynamical fluctuations  affect
the capacity in a more or less subtle manner because the level of dynamical
fluctuations  depends on the overlap.
At zero temperature $(T=0)$ and for $M=1$ neither thermal fluctuations nor
dynamical fluctuations are present. In this limit, perfect matching is
allowed, leading to $\alpha_0=2$ similarly to
Gardner's result~\cite{Gardner87}. However, small departure from $M=1$
induces dynamical fluctuations of the potential of the neurons, so
that the maximum storage capacity $\alpha_0$ decreases rapidly as $M$ is
reduced.  At $T=0$, as shown in Fig.~\ref{fig:a0M},
$\alpha_0$ reaches its maximum at $M=1$ and hence we have the optimal
storage capacity $\alpha_c=2$ regardless  of the tolerance
parameter $m$.
At finite temperatures there always exist thermal  fluctuations, which
prohibit perfect matching. In this case it may be expected that allowing
some error (i.e., $m<1$) increases the capacity. On the other hand,
reducing the overlap introduces dynamical fluctuations and eventually
reduces the capacity if the temperature is not too high ($T<T_1$).
Near $M\approx0$, reduction of  the overlap in general increases the capacity
at any temperature since dynamical fluctuations favor small values of the
overlap.  Here we stress that one should not expect the divergence of the
capacity in the limit $m\rightarrow0$ because the trivial solution
$M_\mu^\ast =0$ is not included.

{}From the curves of $\alpha_0$ it is straightforward to get the optimal
storage
capacity $\alpha_c$ defined by Eq.~(\ref{eq:oscm}) for  given tolerance
parameter and temperature. Typical behavior of $\alpha_c$ as a function of $m$
is shown in Fig.~\ref{fig:acm} at several temperatures.
At temperatures higher than $T_1\,(\approx0.566)$, $\alpha_0$ is a monotonic
decreasing function of $m$. Consequently, we have $\alpha_c(m;T)=\alpha_0(
M=m;T)$, and the curves of $\alpha_c$ are identical to those of $\alpha_0$.
For $T<T_1$, there appears a plateau on which the optimal storage capacity
is constant over some range of the tolerance parameter.
(See Fig.~\ref{fig:acm}. The boundary of this region is displayed by the
dotted line.) For $T<T_2(\approx0.414)$, it is interesting to note that
$\alpha_0$ reaches its maximum near $M\approx1$ and that a large value of the
overlap is mostly favored.

Consider a problem that we want to store and recall $\alpha C$ random patterns
in the network at temperature $T$ at the best performance, that is,
we want the stationary overlap as large as possible.
When $\alpha$ is small, one can easily find a set of couplings
($\{J_{ij}\}$) that yields the stationary value of the overlap near unity.
As $\alpha$ increases, it becomes more difficult to find such a set of
couplings. In general the quality of performance will deteriorate with the
storage ratio $\alpha$.
Since $\alpha_0(M;T)$ is the maximum storage capacity with the
stationary overlap $M$ at temperature $T$, the best performance
$M_p(\alpha;T)$ for given storage ratio $\alpha$ and temperature
$T$ is determined by the largest value of $M$ for which $\alpha_0(M;T)$ is
greater than $\alpha$.  This implies $\alpha = \alpha_c(M_p;T)$ and
the curve of the best performance also corresponds to
the optimal storage capacity.
Therefore Fig.~\ref{fig:acm} also represents curves of the best
performance with
the abscissa and the ordinate denoting $M_p$ and $\alpha$, respectively.
As the number of stored patterns increases, there occurs a first-order
transition from good performance to poor performance at temperatures not
too high ($T<T_1$).
Interestingly, at low temperatures $(T<T_2)$ the network near saturation
naturally favors high-quality performance; there are no networks yielding
low-quality performance.


\section{Analysis using approximate criterion}
         \pindent
In this section, we study the proposed scheme with the approximate
criterion given by Eq.~(\ref{eq:appfpe}) instead of Eq.~(\ref{eq:stfpe})
because it is very difficult to solve the dynamics of neural
networks in general.
In fact the calculation even with the approximation is not easy
and we implement the calculation only for extremely-diluted neural networks.
The validity of the approximation will be tested against the result of
Sec.~III.
The fractional volume $V$ of the space of interactions ($\{J_{ij}\}$)
satisfying Eqs.~(\ref{eq:normal}) and (\ref{eq:appfpe})  is given by
\ban
   V &=& \int[\prod_{i\ne j}dJ_{ij}] \prod_{\mu=1}^{\alpha C}
      \theta\biggl(\frac1N\sum_{i=1}^N\xi_i^\mu\tanh(\frac{\beta
            }{\sqrt{C}}\sum_jJ_{ij}\xi_j^\mu)-m\biggr)
             \prod_i \delta\biggl(\sum_j(J_{ij})^2-C\biggr)  \\
    && \times \biggl\{\int[\prod_{i\ne j}dJ_{ij}]
        \prod_i \delta\biggl(\sum_j(J_{ij})^2-C\biggr)\biggr\}^{-1}\,.
\ean
Although there is no restriction on
the correlations between $J_{ij}$ and $J_{ji}$, different sites $i$ and $j$ are
not decoupled because of Eq.~(\ref{eq:appfpe})\/; thereby it is not easy to
calculate the typical fractional volume $\overline{V} \equiv \exp(\qal \log V
\qar)$, which involves the average $\qal \; \qar$ of $\log V$ over the
distribution of the random patterns $\{\bxi^\mu\}$.
Nevertheless the calculation can be performed for an extremely-diluted network
as discussed in the previous section.
In this case the cumulant expansion can be cut-off as before
at the second order.

Following procedure similar to that in Sec.~III,
we obtain the typical fractional volume in the form
\be
 \overline{V} = \exp\biggl(CN\biggl[\sfrac12\log(1-q)+\sfrac14\log(1-x^2)
    +\sfrac12 \frac{q-rx}{(1-q)(1-x^2)}+\alpha g_2\biggr]\biggr)\,,
     \label{eq:typicalvolume}
\ee
where the function $g_2$ in this case is  given by
\ban
    g_2 &\equiv& \frac1N\int_{-\infty}^\infty\prod_{i=1}^NDt_i\;\log\biggl[
              \int_{-\infty}^\infty\frac{dt_0}{\sqrt{2\pi/N}}
              \exp(-\sfrac12Nt_0^2) \\
           && \times
              \int\prod_{i=1}^NDh_i\;\theta\biggl(\frac1N\sum_{i=1}^N\tanh[
                \beta(\sqrt{1-q}\,h_i-\sqrt{s-r}\,t_0-\sqrt{q}\,t_i)]-m
             \biggr) \biggr]\,.
\ean
with the same notations as in Sec.~III.
Using the integral representation of the $\theta$-function
$$
       \theta(\frac1N\sum_{i=1}^N\tanh h_i-m) = \int_m^1dM\int_{-i\infty}^{
     +i\infty}\frac{d\lambda}{2\pi i/N}\exp\biggl(-N\lambda(M-\frac1N
         \sum_{i=1}^N \tanh h_i)\biggr)
$$
and noting the range of the integral variable $M$, we obtain
$$
     g_2 = -m\lambda-\sfrac12t_0^2+\int Dt\;\log\biggl(\int Dh\;\exp\{\lambda
       \tanh[\beta(\sqrt{1-q}\;h-\sqrt{(1-q)x}\;t_0-\sqrt{q}\;t)]\}\biggr)\,,
$$
where $t_0$ and $\lambda$ are to be determined by the saddle-point equations.
Since $\overline{V}$ depends on $s$ only through $x$, it is straightforward to
show that $\overline{V}$ reaches its maximum at $x=0$ and $t_0=0$.

The optimal storage capacity can be determined according to the condition that
the typical fractional volume shrinks to zero, which happens as $q$ approaches
unity. In this limit, the typical fractional volume given by
Eq.~(\ref{eq:typicalvolume}) has the leading term:
$$
 \overline{V} = \exp\biggl(CN\biggl[\sfrac12\log(1-q)-\alpha m\lambda+
             \frac{\alpha}{1-q}\int Dt\,\Omega_t(H_t)\biggr]\biggr)\,,
$$
where $\Omega_t(H)\equiv -\sfrac12(H+t)^2+\lambda(1-q)\tanh(\beta H)$ in this
case and $H_t$ is again the value of $H$ leading to the maximum of $\Omega_t$.
Note that $H_t$ depends on $\lambda(1-q)$ and $T$ in addition to $t$.
Thus, in the limit $q\rightarrow 1$ and  $\lambda \rightarrow \infty$ with
$\lambda(1-q)$ fixed, the saddle-point equations read
\bal
     \alpha_c &=& \biggl(\int Dt\;\biggl\{t+H_t[\lambda(1-q);T]\biggr\}^2
       \biggr)^{-1} \label{eq:limitalpha} \alp \ealabel{appac}
    m &=& \int Dt\; \tanh\biggl(\beta H_t[\lambda(1-q);T]\biggr)\,,
             \label{eq:limitm}
\eal
where $H_t$ is, by definition, given by the solution of the equation
\be
    \Omega_t'(H) \equiv -(H+t)+\lambda(1-q)\beta[1-\tanh^2(\beta H)] = 0\,.
    \label{eq:ext}
\ee
Equation~(\ref{eq:ext}) has a unique root unless $\lambda(1-q)>(3\sqrt3/4)T^2$
and $t_{-}<t<t_{+}$. [In this range Eq.~(\ref{eq:ext}) has three roots.]
Here $t_{\pm}$ are defined to be
$$
   t_{\pm} \equiv \Omega_{t=0}'[H=-T\tanh^{-1}\{\frac2{\sqrt{3}}\cos(
                  \frac{\pi \pm \phi}{3})\}]
$$
with
$$
        \phi \equiv \cos^{-1}\biggl(\frac{3\sqrt3T^2}{4\lambda(1-q)}\biggr)\,.
$$
At zero temperature it is straightforward to compute $H_t$ and to write the
optimal storage capacity in the form
$$
   \alpha_c = \biggl(\int_0^{\sqrt{2}\erf^{-1}(m)} Dt\,t^2\biggr)^{-1}
$$
while at finite temperatures Eqs.~(\earef{appac})
can be solved numerically.

Figure \ref{fig:am_app} displays the optimal storage capacity $\alpha_c$
as a function of the tolerance parameter $m$ at various temperatures.
The overall dependence of $\alpha_c$ on $m$ is qualitatively different
from that obtained in Sec.~III.
In particular $\alpha_c$ diverges as $m\rightarrow0$. Our
approximate criterion loses its validity near $m=0$ as it should be.
However, $m$-dependence of $\alpha_c$ near $m=1$ is not too disparate from
that of Fig.~\ref{fig:acm} at finite temperatures.
The $\alpha_c$ curves at various temperatures shown in Fig.~\ref{fig:am_app}
are reproduced to expose the detailed behavior near $m=1$ in
Fig.~\ref{fig:com}, which, for comparison, also displays the corresponding
curves obtained in Sec.~III.
At given temperature the two curves indeed coincide with each other in the
limit $m\rightarrow1$. Therefore we conclude that the approximate scheme
based on Eq.~(\ref{eq:appfpe}) is valid for $m$ close to unity.

It is of interest to apply the above scheme to the dynamic model of neural
networks~\cite{Choi88}, where a neuron is forced to have state $s_i=-1$
during the refractory period. As a consequence, Gardner's method cannot be
applicable even at zero temperature.
In the dynamic model, equations describing the time evolution of relevant
physical quantities in general assume the form of differential-difference
equations due to the retardation in interactions. In particular, the activity
$\langle s_i(t) \rangle$ of the \th{i} neuron at time $t$
and the overlap $M_\mu(t)$
between the network state and the \th{\mu} pattern at time $t$ are
determined by the differential-difference equations
\ban
    \frac{d}{dt} \langle s_i(t)\rangle &=&
           (\sfrac12-a)-(\sfrac12+a)\,\langle
       s_i(t)\rangle + \sfrac12(1-\langle s_i(t)\rangle)\,\tanh\biggl(
       \frac{\beta}{\sqrt{N}}\sum_j J_{ij}\langle s_i(t-1)\rangle\biggr) \\
     \frac{d}{dt}  M_\mu(t) &=&
             -(\sfrac12+a)\,M_\mu(t) + \frac1{2N}
           \sum_{i=1}^N\xi_i^\mu(1-\langle s_i(t)\rangle)\,\tanh\biggl(
       \frac{\beta}{\sqrt{N}}\sum_j J_{ij}\langle s_i(t-1)\rangle\biggr)\,,
\ean
respectively. Here $a$ represents the ratio of the refractory period to the
time duration of the action potential.
In the stationary state, the overlap $M_\mu$ takes the form
\be
       M_\mu = \frac{4a}{1+2a}\frac1N\sum_{i=1}^N\frac{\xi_i^\mu \tanh
   \biggl( \frac{\beta}{\sqrt{N}}\ds{\sum_j}J_{ij}\langle s_i\rangle\biggr)
 }{ 1+2a+\tanh \biggl( \frac{\beta}{\sqrt{N}}\ds{\sum_j}J_{ij}\langle s_i
   \rangle \biggr)}.  \label{eq:dyn}
\ee
Since $\xi_i^\mu\sum_jJ_{ij}\xi_j^\mu$ tends to be positive for typical
types of interactions, we may, in the extreme-dilution limit, make an
approximation in Eq.~(\ref{eq:dyn}) as
$$
     M_\mu = \frac1N\sum_{i=1}^N\frac{4a\tanh\biggl( \frac{\beta}{
     \sqrt{N}}\ds{\sum_j}\xi_i^\mu J_{ij}\xi_j^\mu \biggr)}{ (1+2a)^2-\tanh^2
 \biggl( \frac{\beta}{\sqrt{N}}\ds{\sum_j}\xi_i^\mu J_{ij}\xi_j^\mu \biggr)}\,.
$$
The optimal storage capacity of the dynamic model is now determined by
Eqs.~(\earef{appac}) except for that the
$\tanh x$ function is replaced by $4a\,\tanh x/[(1+2a)^2-\tanh^2x]$.
Unlike the Hopfield model which discretizes the time, the dynamic model
takes into account the existence of relevant time scales, and consequently,
displays the overlap $M_\mu=1/(1+a)$ in the case of perfect recalling.
This is reflected in the equation corresponding to Eq.~(\ref{eq:limitm}).
For comparison with the Hopfield model, therefore, we rescale the
tolerance parameter $m$ by $\widetilde{m}\equiv(1+a)m$, and finally get
$$
   \alpha_c(\widetilde{m}) =
       \biggl(\int_0^{\sqrt{2}\,\erf^{-1}(\widetilde{m})}Dt\;t^2\biggr)^{-1}
$$
at zero temperature.


\section{Discussion}
                   \pindent
We have proposed a new method to study the optimal storage property
of neural networks at finite temperatures and investigated the optimal storage
capacity $\alpha_c$ for an extremely-diluted network,
as a function of temperature $T$ and the tolerance parameter $m$.
At zero temperature, it has been found that $\alpha_c=2$ regardless of
the tolerance parameter whereas at finite temperature $\alpha_c$ vanishes
in the perfect matching limit ($m\rightarrow 1$), in general increasing with
the tolerance.
The best performance for given storage ratio $\alpha$ and temperature has
been also obtained. At low temperatures $(T<T_1\approx0.566)$ the network
exhibits a first-order transition from high-quality performance to
low-quality performance as the number of stored random patterns is increased.
High-quality performance seems to be naturally favored by
extremely-diluted networks if the level of noise is not so high.
We have also studied an approximate scheme, which yields qualitatively
different results except near $m=1$.  The crude approximation
$\langle s_i \rangle\approx\xi_i^\mu$ used in Eq.~(\ref{eq:appfpe}) has been
found not good in the whole range of $m$; for $m$ close to unity, however,
it can be regarded as an accurate approximation.

Instead of the proposed criterion for recalling, one may consider a slightly
different criterion: The time average of $\xi_i^\mu s_i(t)$ {\it for each
site i\/} should be greater than $m$. In the same spirit as in
Eq.~(\ref{eq:appfpe}), one may consider the  problem of
calculating the typical fractional volume of the space of interactions
satisfying Eq.~(\ref{eq:normal}) and
\be
    \xi_i^\mu \tanh(\frac{\beta}{\sqrt{N}}\sum_j J_{ij}\xi_j^\mu)\;>m
    \label{eq:third}
\ee
for each $i$. The problem is then equivalent to Gardner's problem with her
parameter $\kappa$ given by $\kappa=T\tanh^{-1}m$, which implies that
the required basin of attraction grows larger with the level of synaptic
noise and with the accuracy of recalling.
At zero temperature this leads to the optimal capacity $\alpha_c=2$
regardless of $m$,
whereas at finite temperatures $\alpha_c$ increases with the tolerance.
Despite their resemblance,
the behavior of the optimal storage capacity with the criterion
(\ref{eq:third}) is qualitatively different from that of (\ref{eq:appfpe}).
Although the argument of tanh in both cases is a sum over many sites $j$,
it should be strongly correlated with $\xi^\mu_i$ if the network ($\{
J_{i j} \}$) is to function as associative memory. In general, the overlap
on site $i$, given by the r.h.s.\ of Eq.~(\ref{eq:third}), will vary from
site to site according to some distribution with finite variance. The
criterion (\ref{eq:appfpe}) in the large $N$ limit requires the
overlap averaged over that distribution be  greater than $m$ while that of
(\ref{eq:third}) demands that the overlap be greater than $m$ for each site.

Finally, there are several points for further investigation.
Since we have assumed the replica- and site-symmetric ansatz in the
calculation of the typical fractional volume,
its stability against replica-symmetry-breaking should be checked.
It is also of interest to extend our results to the fully-connected
networks and other types of networks.

\ \\
{\Large \bf Acknowledgement}
\\
This work was supported in part by the Korea Science and Engineering
Foundation through the Center for Theoretical Physics,
Seoul National University.


\newpage
\centerline{\large \bf Figure Captions}
\vspace{0.41in}

\begin{list}{{\large \bf Fig. \arabic{sen}}$\mbox{\ }$}{\usecounter{sen}
             \setlength{\leftmargin}{\rightmargin}}
\item Typical behavior of the maximum storage capacity $\alpha_0$ as a
      function of the stationary overlap $M^\ast$ at various temperatures:
      From top to bottom, $T=0,\;0.3,\;0.5$, and 0.7, respectively.
      Detailed behavior at $T=0.3$ and 0.5 is displayed in the inset.
      \label{fig:a0M}
\item Typical behavior of the optimal storage capacity $\alpha_c$
       as a function of the tolerance parameter $m$ at various temperatures:
       From top to bottom, $T=0.3,\;0.5$, and 0.7, respectively. The dotted
       curve shows the boundary of the region in which $\alpha_c$ is constant.
      \label{fig:acm}
\item The optimal storage capacity $\alpha_c$ as a function of  the tolerance
      parameter $m$ at $T=0,\;0.3,\;0.5,\;0.7$, and 2.0 when the alternative
      approximate criterion given by Eq.~(\ref{eq:appfpe}) is used.
    \label{fig:am_app}
\item Detailed behavior of the optimal storage capacity $\alpha_c$ for the
      tolerance parameter $m$ near unity. Solid lines and dotted lines are
      results of Eqs.~(\ref{eq:stfpe}) and (\ref{eq:appfpe}), respectively.
    \label{fig:com}
\end{list}

\end{document}